\documentclass[aps,prl,showpacs,showkeys,twocolumn]{revtex4-1}
\usepackage{graphicx,hyperref,amsmath,amsfonts}
\usepackage{epstopdf,color,bm,multirow,rotating,soul}

\begin{document}
\title{RCF-defense for quantum randomness}
\author{N.S. Perminov$^{1,2,*}$, O.I. Bannik$^{1}$, D.Yu. Tarankova$^{3}$ and R.R. Nigmatullin$^{4}$}
\affiliation{$^{1}$Kazan Quantum Center, Kazan National Research Technical University n.a. A.N.Tupolev-KAI, 10 K. Marx, Kazan 420111, Russia}
\affiliation{$^{2}$Zavoisky Physical-Technical Institute, SSS FRC KSC RAS, 10/7 Sibirsky Tract, Kazan 420029, Russia}
\affiliation{$^{3}$Institute of Radio-Electronics and Telecommunications, Kazan National Research Technical University n.a. A.N.Tupolev-KAI, 10 K. Marx, Kazan 420111, Russia}
\affiliation{$^{4}$Department of Radio-Electronics and Information-Measuring Technique, Kazan National Research Technical University n.a. A.N.Tupolev-KAI, 10 K. Marx, Kazan 420111, Russia}
\email{qm.kzn@ya.ru}
\pacs{ 03.67.-a, 06.20.Dk, 05.40.−a.}
\keywords{quantum information, ranged correlation functions, RCF-defense, quantum randomness, physical random number generator.}

\begin{abstract}						
Authors develop new nonparametric methods for verification and monitoring for quantum randomness based on the ranged correlation function (RCF) and a sequence of the ranged amplitudes (SRA). We carried out RCF-analysis of different topology subsamples from raw data of the prototype of a quantum random number generator on homodyne detection. It is shown that in the real system there are weak local regression relations, for which it is possible to introduce a robust criterion of significance, and also precise SRA-identification of long samples statistics is made. The obtained results extend the traditional entropy methods of the useful randomness analysis and open the way for creation of new strict quality quantum standards and defense for physical random numbers generators.
\end{abstract}


\maketitle

\textit{Introduction.}
Quantum randomness as a phenomenon within modern quantum physics \cite{Pusey2012} and mathematical logic \cite{Paterek2010} has a very special status, which makes it possible to attribute quantum randomness to the class of effects in partially deterministic complex systems \cite{Nigmatullin2015}. In addition, in experimental physics there is no ideal measuring technique capable of directly measuring the “ideal” quantum randomness. Therefore, the development of stable methods for nonparametric analysis of quantum time series \cite{Smirnov2018,Perminov2018} is necessary for the development of the fundamental criteria for the quantitative parametrization of quantum randomness.

In quantum communications \cite{Gisin2002} and the problems of generating true random numbers \cite{Balygin2017,Gabriel2010}, the question of the quantitative assessment of the quality of the initial raw randomness is one of the central ones. Common methods for analyzing raw randomness in quantum systems often come down to identifying the empirical frequencies of long samples and calculating the autocorrelation function, but poorly predict methods for local analysis \cite{Box2015}. Therefore, we believe that new precision methods for the correlation analysis of time series \cite{Nigmatullin2006} will be able to expand the possibilities of error correction in quantum informatics at the post-processing stage, as well as will make it possible to enhance the secrecy and defense of communications at the physical level.

In this paper, we develop new non-parametric methods for verifying and monitoring quantum chance based on the ranged correlation functions (RCF) and its analogues using a sequence of ranged amplitudes (SRA). An RCF-analysis of various topology subsamples from the raw data of the prototype quantum generator of random numbers on homodyne detection was performed. For the studied series, a method of precision SRA-identification of statistics of short samples was proposed, and weak local regression relations were found, for which stable criteria of significance were introduced. The results obtained significantly expand the traditional entropy analysis methods and open the way for the development of common quantum standards and defense for physical randomness.

\textit{Ranging.}
Any time series $\{x_k\}$ ($k=\overline{1,N}$), consisting of real or complex numbers, can be ranged according to the chosen measure by decreasing (or increasing) and get a sequence of ranged amplitudes (SRA) of the form $\{x_n\}$ ($n=\overline{1,N}$), where the index $n$ is the index in SRA \cite{Nigmatullin2006}.
Obviously, according to this definition, the sequence of SRA $\{x_n\}$ is composed of exactly the same elements as the original sequence $\{x_k\}$; therefore, SRA is a non-invasive (without loss of information) statistical quantitative characteristic of a data sample \cite{Nigmatullin2006}. SRA is related to the distribution function by the following approximate relation (where $N$ is the sample size) \cite{Smirnov2018,Perminov2018,Stephens1992,Kamps1995}:
\begin{align}\label{eq_1}
& F(x_n;N)=(N+1-n(x_n))/N.
\end{align}
Note also that any (even non-smooth and infinite) statistical functions (statistical averages over the initial sample) of a given sample $\{x_k\}$ and SRA $\{x_n\}$ strictly coincide.
Mathematically, this can be written as a condition
$G[\{x_k\}]=\sum_kG(x_k)=\sum_nG(x_n)=G[\{x_n\}]$
for any function $G(x)$ (for example, $G(x)$ it can be some entropy measure). Therefore, any entropic measure and sum-functions of the sample (including the SRA) on the information capacity equivalent or superior to the original SRA. In this sense, SRA-analysis significantly expands the possibilities of entropy analysis.

Many of the generalized correlation functions \cite{Nigmatullin2006} of a pair of samples $\{x_k\}$, $\{y_k\}$ are in fact the prototype of three SRA: SRA $\{x_k\}$, SRA $\{y_k\}$ and SRA of the direct product of samples $\{w_n^2=(x_ky_k)_n\}$. We will consider this triple as the base of the ranged correlation functions (RCF). In the future, for all samples by default, we will use the normalized scale, that is $\{x_k\rightarrow(x_k-min(x_k))/(max(x_k)-min(x_k))\}$. Such a normalization is the only mapping that does not destroy the structure of linear regression relations, allows to correctly define the generalized correlation functions \cite{Nigmatullin2006} on the domain of complex variables, which is necessary to identify nonlinear regressions \cite{Nigmatullin2015,Box2015,Nigmatullin2006,Kamps1995}. Normalization also allows you to select only one sample $\{w_n\}$ from among the whole three described SRA to verify redundant correlations based on RCF.

\textit{RCF-analysis of subsamples.}
Quantitative analysis of internal correlations and randomness can be most effectively implemented on the basis of significantly different topology subsamples of the initial sample to which the quality criteria should be presented. For the original sample $x_j$ of size $2N$ ($j=\overline{1,2N}$), we selected 4 subsets of the same size $N$: $x_{1,k}=x_k$, $x_{2,k}=x_{k+N}$, $x_{3,k}=x_{2k-1}$, $x_{4,k}=x_{2k}$ ($k=\overline{1,N}$). Note that the properties of these 4 subsamples are homogeneous with respect to time, and the samples 1,2 and 3,4 form disjoint covers of the initial sample with different topology.
Therefore, it is possible to consider two pairs independently: 1\&2 and 3\&4. It is obvious that due to the construction of the pair 1\&2 it is able to feel ultra-long correlations (typical, for example, for mathematical generators of pseudo-random numbers), and the pair 3\&4 is able to “see” local linear regression connections, expanding the possibilities of autocorrelation function for the initial sample of length $2N$.

The motivation for the in-depth study of this four subsamples was a statistically significant (by an order of magnitude) experimental observation of the difference in Pearson correlation coefficients in the pair 1\&2 -- $R_{12}^2=6,9\cdot10^{-4}$ and pair 3\&4 -- $R_{34}^2=8,4\cdot10^{-3}$ for raw data of size $N=10^6$ obtained on the prototype of the quantum random number generator on homodyne detection. Discovered a local regression to the challenge of sustainable criteria of significance for nonrandomness in the source sample. The solution of this problem is proposed to make on the basis of methods of SRA-identification \cite{Nigmatullin2015} statistics, which, as we have shown earlier, can be effectively applied even to short samples of quantum data \cite{Smirnov2018,Perminov2018}.

\textit{W-statistics of the product of samples.}
Traditional methods of analysis of raw quantum randomness are often reduced to the identification of empirical frequencies or their histograms for long samples \cite{Balygin2017,Gabriel2010}, which introduces certain identification errors associated with the invasiveness of these methods. For raw data obtained as a result of homodyne detection of quantum randomness \cite{Gabriel2010}, it is considered correct to obtain normally distributed empirical frequencies. However, a natural question immediately arises about the distribution of the cumulative frequency function (discrete integral of empirical frequencies) and, accordingly, the SRA-distribution associated with it (\ref{eq_1}). The cumulative distribution function can be parameterized both by the error function (integral of the normal distribution) and by the sum of two normal distributions following from the discrete integration of the normal distribution.
This fundamental aspect of ambiguity is related to the discretization of the data, which leaves the possibility to use empirical frequencies, cumulative frequencies or SRA to identify statistics. But the traditional criteria for the significance of theoretical models in statistics \cite{Stephens1992} have been proved for SRA, so we tend to the version that for the most correct identification of statistics, the normality test should be understood as the proximity of the fitting function for the inverse of the SRA function (see (\ref{eq_1})) to the error function $n(x)=A+B\cdot erf((x-x_0)/dx)$. Our calculations have shown that the accuracy of the normalized SRA fitting due to the error function $n(x)=(1+erf((x-x_0)/dx))/2$ is higher than the accuracy of the parameterization of the empirical frequency distribution by the Gaussian normal distribution. Therefore, we will further use the normality conjecture in the sense of the error function for other ranged data as well.

The main mathematical task of the traditional correlation statistical analysis of a pair of samples $\{x_k\}$, $\{y_k\}$ is to identify the symmetric relations given by sum-functions $G[\{x_k;y_k\}]=G[\{y_k;x_k\}]$ (in the spirit of entropy analysis), which can be written directly in the basis of symmetric functions $G=G_{sym}[\{w_k^2=x_ky_k;r_k^2=x_k^2+y_k^2\}]$ or angular symmetric functions (spherical coordinates) $G_{angle}[\{\varphi_k=\textrm{arcsin}(2w_k^2/r_k^2)/2;r_k^2\}]$. In the basis of symmetric functions, we usually further consider the variable-split sum-functions, what leads us in the analysis of $G_{sym}$ to an independent consideration of the series $\{w_k\}$ and $\{r_k\}$, and hence their SRA $\{w_n\}$ and $\{r_n\}$.
In this paper, we restrict ourselves to the consideration of SRA $\{w_n\}$, built on the normalized series $\{x_k\}$ and $\{y_k\}$. It is important to note that the variance of such w-statistics of the product of samples (SRA $\{w_n\}$) in its construction can be associated with the standard Pearson match criterion $R^2$, greatly expanding its capabilities.

\begin{figure}[h]
\includegraphics[width=0.45\textwidth]{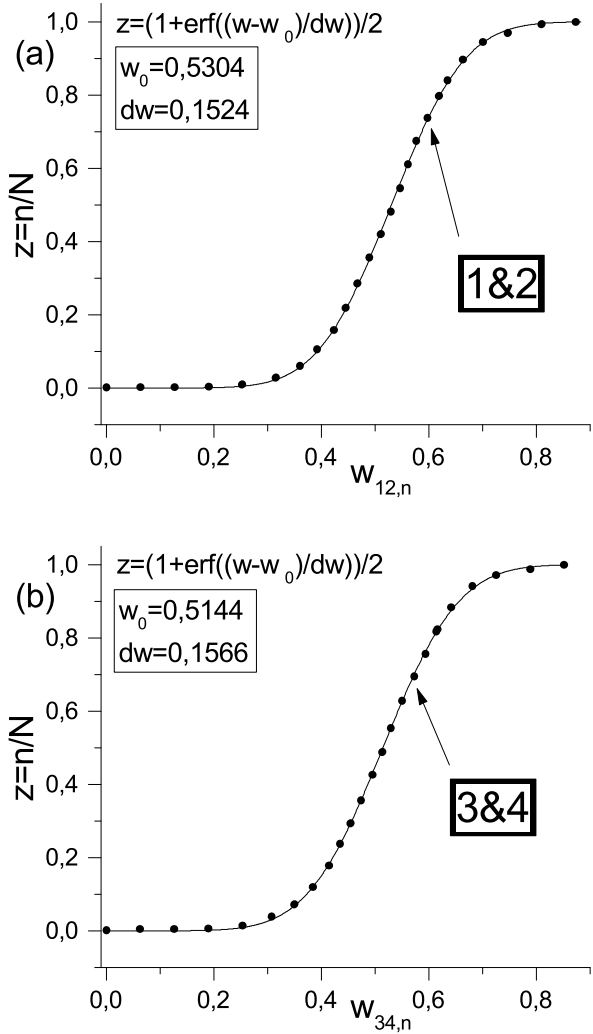}
\caption{W-statistics of the product of topologically different subsamples $\{w_{12,n};z_n=n/N\}$ -- (a) and $\{w_{34,n};z_n=n/N\}$ -- (b).}
\label{fig_1}
\end{figure}

Previously, we found the difference between $R^2_{12}=6,9\cdot10^{-4}$ and $R^2_{34}=8,4\cdot10^{-3}$ for two different subsamples from the raw data of the prototype of a quantum random number generator on homodyne detection. To see the difference between subsamples 1\&2 and 3\&4 at the level of w-statistic and to demonstrate the sensitivity of RCF-analysis, we built and parametrized at the expense of the error function $z=(1+erf((w-w_0)/dw))/2$ sets $\{z_n=n/N;w_{12,n}\}$ and $\{z_n=n/N;w_{34,n}\}$. For depicted in Fig.\ref{fig_1} curves fitting accuracy was approximately $0.9993$, and the obtained parameters of fitting $\{w_0=0,5304;dw=0,1524\}$ for 1\&2 and $\{w_0=0,5144;dw=0,1566\}$ for 3\&4 indicate a statistically significant difference between the two pairs of subsamples.

The criterion for the significance of correlations in this situation is not one Pearson parameter $R^2$, and 2 parameters $\{w_0;dw\}$ of the model error function. The sensitivity of the RCF-analysis technique can also be increased by improving the model fitting function \cite{Nigmatullin2015,Nigmatullin2006} by introducing an additional degree of freedom $\theta$ in the form of an additional non-extensiveness parameter to the error function of the $z=A+B\cdot erf( (w-w_0)^{\theta}/dw)$ type, which corresponds to the availability of effective memory in time series. However, taking into account additional parameters of fitting with the help of standard methods of mathematical analysis is a nontrivial task in many practical situations and the question of the implementation of algorithms for accounting non-extensiveness in statistical distributions requires a separate in-depth study \cite{Nigmatullin2015,Nigmatullin2006}.

\textit{Angle analysis of randomness.}
An additional method for the correlation analysis of randomness on the basis of symmetric functions is to consider the distributions of the angles $\{\varphi_{k}=\textrm{arcsin}(2w_k^2/r_k^2)/2\}$ and the radii $\{r_k=(x_k^2+y_k^2)^{1/2}\}$, constructed from the centered data $\{x_k\rightarrow x_k-\langle x_k\rangle\}$. The SRA $\{r_n;z=n/N\}$ distribution is normal in the sense of the error function with a fit accuracy of about $0.9999$, and the $\{\varphi_n;z=n/N\}$ angle distribution has the character of a uniform distribution ($\varphi_n\cong n/N$) with about the same degree of accuracy.
Therefore, to identify potential regression links, we use a more subtle criterion based on the SRA-analysis of discrete derivatives of the initial distributions $\{\varphi'_{k}=\varphi_{k+1}-\varphi_{k}\}$ and $\{r'_{k}=r_{k+1}-r_{k}\}$ characterizing the heterogeneity of the angular distributions.
\begin{figure}[h]
\includegraphics[width=0.45\textwidth]{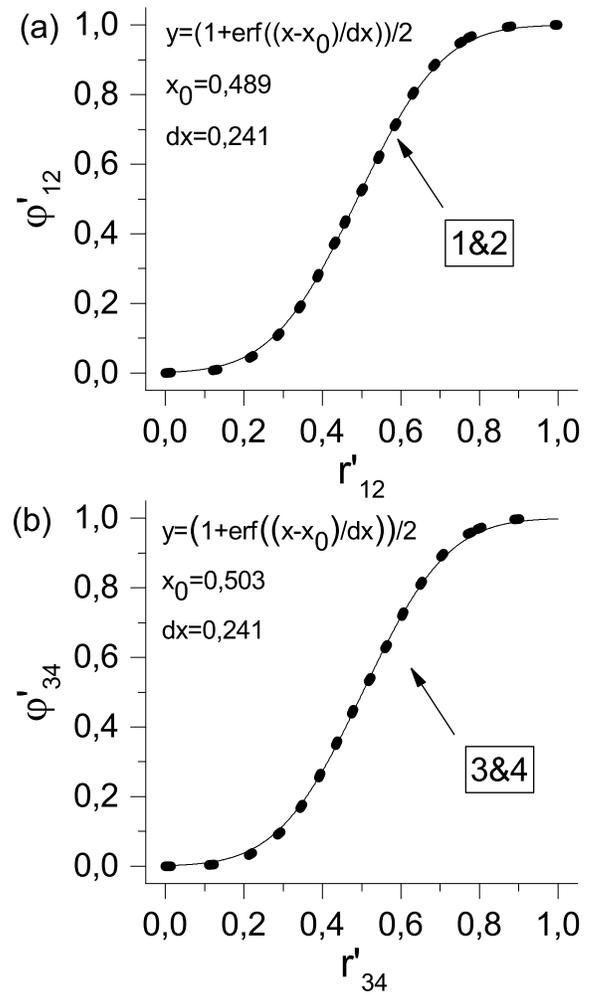}
\caption{The distribution of the inhomogeneity of the angular variables for 1\&2-subsamples $\{r'_{12,n};\varphi'_{12,n}\}$ -- (a) and 3\&4-subsamples $\{r'_{34,n};\varphi'_{34,n}\}$ -- (b).}
\label{fig_2}
\end{figure}
In Fig.\ref{fig_2}, for illustration, we constructed relative dependences of the velocities of the radius-angle $\{r'_{12,n};\varphi'_{12,n}\}$ and $\{r'_{34,n};\varphi'_{34,n}\}$ for two subsamples 1\&2 and 3\&4 of different topology (size $N=10^6$), for which the structure of the normal distribution was obtained (fitting accuracy $0,9995$) with differing adjustable parameters.

Since the distribution of angles for both pairs 1\&2 and 3\&4 has the structure of a uniform random variable, it was possible to represent the series $\{\varphi_k;r_k\}$ in a split form $\{\textrm{sign}(\varphi'_k);\textrm{abs}(\varphi'_k);r'_k\}$, where randomness analysis can be carried out simultaneously for all three components. At the same time, the series $\{\textrm{sign}(\varphi'_k)\}$ for both 1\&2 and 3\&4 contains an extremely small fraction of zeros ($\sim2,5\cdot10^{-4}\pm3\cdot10^{-6}$) and consists of numbers $\{-1;1\}$, which, after replacing $-1\rightarrow 0$, give us a bit sequence. For these sequences, the standard set of NIST cryptographic tests \cite{Rukhin2001} can be applied for two different cases and a $p-value$ criterion of significance is obtained ($p-value>0,01$ indicates the test has passed).

\begin{figure}[h]
\includegraphics[width=0.45\textwidth]{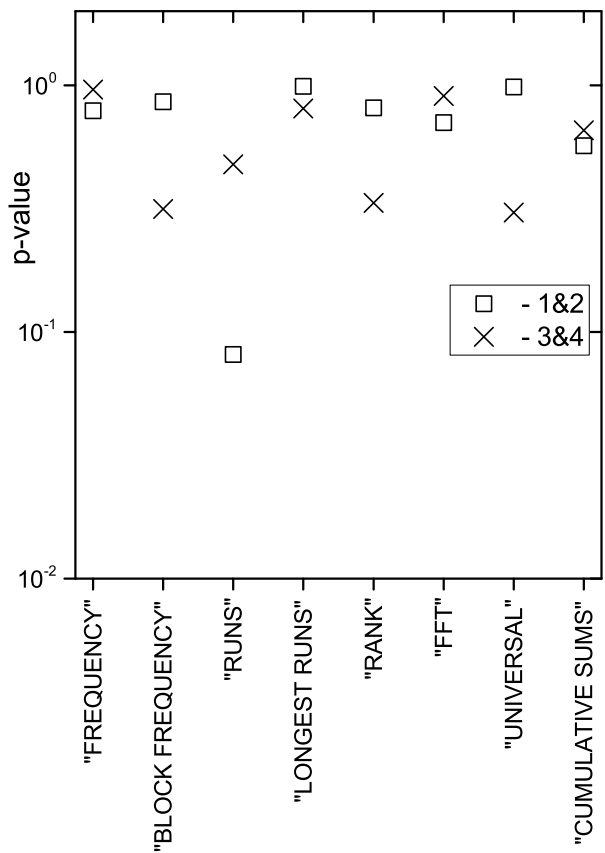}
\caption{Basic NIST-tests \cite{Rukhin2001} for split quantum randomness, given by the inhomogeneities of the angular distributions $\{\textrm{sign}(\varphi'_{12,k})\}$ and $\{\textrm{sign}(\varphi'_{34,k})\}$.}
\label{fig_3}
\end{figure}

We used subsamples of length $N=10^6$ and found that a part of quantum randomness split off in such a way satisfies the cryptographic criteria of a true random bits, which can be immediately used in quantum cryptography applications \cite{Gisin2002,Gabriel2010}. The results of passing the main tests of NIST \cite{Rukhin2001} are shown in Fig.\ref{fig_3}.

For the presented NIST-tests, the standard notation was used: "FREQUENCY" -- frequency test, "BLOCK FREQUENCY" -- frequency test in blocks, "RUNS" -- check the "holes", “LONGEST RUNS” - check the "holes" in the subsamples, “RANK" -- check of matrix ranks, "FFT" -- spectral test, "UNIVERSAL" -- Maurer's universal statistical test, "CUMULATIVE SUMS" -- check of cumulative sums.
Both sets of binary numbers $\{\textrm{sign}(\varphi'_{12,k})\}$ for subsamples 1\&2 and $\{\textrm{sign}(\varphi'_{34,k})\}$ for subsamples 3\&4, as shown by additional studies, successfully pass all randomness tests. Thus, we managed to present the initial randomness through its subsamples in such a three-component form $\{\textrm{sign}(\varphi'_k);\textrm{abs}(\varphi'_k);r'_k\}$ that the first component is a true random variable, as verified by NIST tests, and to the second and third components it is possible (in parallel with the first component) apply the criteria based on the SRA and RCF methods. These circumstances are important for the implementation of an effective procedure for extracting the final binary randomness from the raw data of a physical random number generator, which will have a regulated structure with the possibility of reliable statistical monitoring of internal security parameters and self-protection of the physical generator of random numbers.

\textit{Conclusion.}
The development of general criteria for the quality of randomness in view of the absence of regression equations that clearly distinguish non-randomness remains a task for the future. But now we can present stable multi-parameter intermediate criteria based on the parameterization of the SRA curves, the product of the samples and the SRA distributions of the angular variables of the subsamples of different topology, which extended the traditional methods of analyzing quantum randomness. A significant advantage of composite angular analysis with splitting of a species $\{\textrm{sign}(\varphi'_k);\textrm{abs}(\varphi'_k);r'_k\}$ is the possibility of accurately separating high-quality binary randomness $\{\textrm{sign}(\varphi'_k)\}$ from the initial data set and verifying it using the standard set of NIST testing methods \cite{Rukhin2001}.

Due to non-invasiveness, the method of SRA and RCF can also be applied in the area of identification of various signal sources \cite{Nigmatullin2015,Spitsyn2016,Umnov2016} and noise \cite{Nigmatullin2006}. The demonstrated advantages of the ranged analysis open up new possibilities for a quantitative description of the quality of useful quantum randomness and the introduction of universal quantum standards and defense in the area of security of optical and quantum communications.

\textit{Acknowledgments.}
Research of noise in the area of photonics and quantum technologies is financially supported by a grant of the Government of the Russian Federation, project No. 14.Z50.31.0040, February 17, 2017 (experimental part). The work is also partially financially supported by the grant of young scientists of RT No 06-36-ts-G 2018 (theoretical part).

\bibliographystyle{apsrev4-1}
\bibliography{RCF_DQR}

\end{document}